\documentclass[usenatbib]{mn2e}
\usepackage{amsfonts}
\usepackage{amsmath}
\usepackage{graphicx}
\usepackage{natbib}

\def\mnras{Mon.\ Not.\ R.\ Astron.\ Soc.}

\def\prd{Phys.\ Rev.\ D}

\title[Effective cosmology and excursion set equivalence]
      {On the equivalence between the effective cosmology and 
       excursion set treatments of environment}

\author[M. C. Martino \& R. K. Sheth]
       {Matthew C. Martino \& Ravi K. Sheth\thanks{E-mail:  shethrk@physics.upenn.edu}\\
 Department of Physics \& Astronomy, University of Pennsylvania, 
 209 S. 33rd Street, Philadelphia, PA 19104, USA}

\def\ltsima{$\; \buildrel < \over \sim \;$}
\def\lsim{\lower.5ex\hbox{\ltsima}}
\def\gtsima{$\; \buildrel > \over \sim \;$}
\def\gsim{\lower.5ex\hbox{\gtsima}}

\begin{document}
\pagerange{\pageref{firstpage}--\pageref{lastpage}}

\maketitle

\label{firstpage}

\begin{abstract}
 In studies of the environmental dependence of structure formation, 
 the large scale environment is often thought of as providing an 
 effective background cosmology:  e.g. the formation of structure in 
 voids is expected to be just like that in a less dense universe 
 with appropriately modified Hubble and cosmological constants.  
 However, in the excursion set description of structure formation 
 which is commonly used to model this effect, no explicit mention 
 is made of the effective cosmology.  Rather, this approach uses the 
 spherical evolution model to compute an effective linear theory 
 growth factor, which is then used to predict the growth and evolution 
 of nonlinear structures.  
 We show that these approaches are, in fact, equivalent:  a 
 consequence of Birkhoff's theorem.  We speculate that this 
 equivalence will not survive in models where the gravitational 
 force law is modified from an inverse square, potentially making 
 the environmental dependence of clustering a good test of such 
 models.  
\end{abstract}


\section{Introduction}
One of the standard predictions of nonlinear hierarchical structure 
formation models is the abundance of virialized structures 
\cite{ps74, st99, virgo01}.  
Simulations show that this abundance depends on the large scale 
environment:  the ratio of massive to low mass objects is larger in 
dense regions \cite[e.g.,][]{fwde88}.  Recent measurements in 
galaxy surveys appear to bear this out:  the virial radii of objects 
in underdense regions are smaller, consistent with their having 
smaller masses \cite[e.g.,][]{as07}.

This paper is motivated by the fact that there are currently in the 
literature three methods for estimating how the mass function of 
virialized halos depends on the environment which surrounds them.  
The first, and perhaps easist to implement, is based on the excursion 
set approach \cite{mw96, st02}.  The second argues that halos which 
form in, say, voids should be thought of as forming in a less dense 
background cosmology, so the mass function is that in a universe with 
 $\Omega_{\rm void} = \Omega_0(1+\Delta_{\rm void})$
\cite[e.g.,][]{glkh03}.  The third is similar, but notes that to 
correctly estimate the background cosmology, one must account not 
only for the lower density in a void, but for the fact the effective 
Hubble constant of the void cosmology is larger than in the background 
\cite[e.g.,][]{gv04}.  One way of thinking about the effective Hubble 
constant is that it ensures that the effective cosmology has the same 
age as the background cosmology. (The cosmological constant is, 
of course, constant, but when expressed in units of the critical 
density in the effective model, it is modified because the critical 
density depends on the effective Hubble constant.)
In Section~\ref{equivalence}, we use the spherical evolution model 
to show that the first and third methods are equivalent 
\cite[although][state otherwise]{gv04}, and that both are 
incompatible with the second method (which incorrectly ignored 
the change to the Hubble constant).  

There has been recent interest in the fact that the formation 
histories of halos of fixed mass depend on their environment 
\cite{st04, gao}, an effect which is not predicted by the simplest 
excursion set methods \cite[e.g.,][]{w96}.
So one might have wondered if this is where the difference between 
the excursion set approach and one based on the effective cosmology
is manifest.  In Section~\ref{evolution} we show that in this case 
also, the two approaches are equivalent.

A final section summarizes our results, and speculates that the 
equivalence we have shown will not survive in models models where 
the force law has been modified from an inverse square.


\section{The calculation}\label{equivalence}
The main point of the following calculation is to show explicitly 
that, at least for cosmologies with no cosmological constant, 
the environmental dependence of halo abundances can be described 
using the excursion set approach \cite[e.g.,][]{mw96, st02}.  
Namely, one need not worry about the details of the effective 
cosmology associated with the region surrounding the perturbation 
\cite[as do][]{gv04}; it is enough to compute an effective growth 
factor using the spherical collapse model.  
Although we have phrased our discussion in terms of an $\Omega_0=1$ 
background cosmology, it is obviously applicable to arbitrary 
values of $\Omega_0$.  
Our analysis suggests that this remains true when the background 
cosmology has $\Lambda\ne 0$.  

For what follows, it is useful to recall that the age-redshift 
relation in an $\Omega_0=1$ cosmology is given by 
$H(z)t(z) = 2/3$, where $H$ is the Hubble constant.  
In an open universe, this relation is 
\begin{equation}
 H(z)t(z) = \frac{1}{1-\Omega(z)} -\frac{ \Omega(z)/2}{(1-\Omega(z))^{3/2}}
            \cosh^{-1} \left({2\over \Omega(z)}-1\right)
 \label{agez}
\end{equation}
where 
\begin{equation}
 \Omega(t) = \frac{\Omega_0/a(t)^3}
                 {\Omega_0/a(t)^3 + (1-\Omega_0)/a(t)^2},
 \label{Omegat}
\end{equation}
with the convention that $a(t_0)\equiv 1$, so $a(t) \equiv (1+z)^{-1}$, 
and $\Omega_0\equiv \Omega(t_0)$ \cite{peebles80}.

The linear theory growth factor is $D(t)=a(t)$ if $\Omega_0=1$, 
and if $\Omega_0<1$ then  
\begin{equation}
 D(t) =\frac{5 \Omega_0/2}{(1-\Omega_0)}\left( 1 + \frac{3}{x} + 3 \sqrt\frac{1+x}{x^3}\,\ln\left[\sqrt{1+x} - \sqrt{x}\right]\right)
\end{equation}
where $x=a(t)\, (1-\Omega_0)/\Omega_0$ \cite{peebles80}.

\subsection{The spherical evolution model}
The spherical evolution model describes the evolution of the size 
$R$ of a spherical region in an expanding universe:  
\begin{equation}
 {{\rm d}^2R\over {\rm d}t^2} = -{GM(<R)\over R^2}.
\end{equation}
It provides a parametric relation between the density contrast 
predicted by linear theory 
 $\delta(t_0) = D(t_0)/D(t_{\rm init})\,\delta(t_{\rm init})$, 
the nonlinear overdensity $\Delta$, and the infall speeds $v_{\rm pec}$ 
\cite{gg72, ps80, peebles80, paddy93, pt02}.
Here $D(t_0)$ is the linear theory growth factor at time $t_0$, 
and we will often use the shorthand,
 $\delta_0 = \delta(t_0)$.  

If $\Omega=1$, then 
\begin{eqnarray}
 {M\over 4\pi R^3\bar\rho/3} &\equiv& 1+\Delta = f(\theta),
 \qquad {v_{\rm pec}\over HR} \equiv g(\theta), 
 \qquad {\rm and}\nonumber\\
 \delta_0 &\equiv& h(\theta),
 \label{scmodel1}
\end{eqnarray}
where 
\begin{eqnarray*}
 f(\theta) &=& \left\{ \begin{array}{ll}
   (9/2)\,(\theta - \sin\theta)^2/(1-\cos\theta)^3 \\
   (9/2)\,(\sinh\theta - \theta)^2/(\cosh\theta - 1)^3 \end{array}\right. ,\\
 g(\theta) &=& \left\{ \begin{array}{ll}
            (3/2)\,\sin\theta\,(\theta - \sin\theta)/(1-\cos\theta)^2 \\
            (3/2)\,\sinh\theta\,(\sinh\theta - \theta)/(\cosh\theta - 1)^2 
          \end{array}\right. ,\\
 h(\theta) &=& \left\{ \begin{array}{ll}
 (3/5)\,(3/4)^{2/3}\,(\theta-\sin\theta)^{2/3} \\
 -(3/5)\,(3/4)^{2/3}\,(\sinh\theta-\theta)^{2/3} \end{array}\right. ,
\end{eqnarray*}
where the first expression in each pair is for initially overdense  
perturbations and the second is for underdense ones.  
Overdense perturbations eventually collapse, the final collapse 
being associated with the value $\theta=2\pi$, at which time the 
linear theory density is 
 $\delta_{\rm c1} \equiv (3/5)(6\pi/4)^{2/3} = 1.68647$.  
In this section, we use the subscript 1 to indicate that this value 
is associated with $\Omega_0=1$.

If $\Omega_0<1$, then only perturbations above some density 
$\delta_{\rm min}$ will collapse, and 
\begin{eqnarray}
 1+\Delta &=& {f(\theta)\over f(\omega)},
 \qquad 
 {v_{\rm pec}\over H_{\rm \omega}R} = {g(\theta)\over g(\omega)} - 1,
 \qquad {\rm and}\nonumber\\
 {\delta_0\over\delta_{\rm min}} &=&- {h(\theta)\over h(\omega)} + 1,
 \label{scmodel}
\end{eqnarray}
where 
\begin{eqnarray}
 \omega = \left\{ \begin{array}{ll}
       \arccos (2/\Omega_0 - 1) & {\rm if\ closed}\\
       {\rm arccosh} (2/\Omega_0 - 1) & {\rm if\ open}
     \end{array}\right. ,
 \label{Uomega}
\end{eqnarray}
$H_{\rm \omega}$ is the Hubble constant, and
\begin{equation}
 \delta_{\rm min} = {9\over 2} {\sinh\omega\, (\sinh\omega - \omega)\over
                            (\cosh\omega -1)^2} - 3.
 \label{special}
\end{equation}
Complete collapse is again associated with $\theta=2\pi$, and we will 
write the critical linear density required for collapse as 
\begin{equation}
 \delta_{{\rm c}\omega} = \delta_{\rm min}\,[1 - \delta_{\rm c1}/h(\omega) ].
\end{equation}
It happens that $\delta_{\rm c\omega}$ 
depends only weakly on $\Omega_0$.  When $\Omega\to 1$, then 
$\delta_{\rm c\omega} \to (3/5)(6\pi/4)^{2/3} = 1.68647$, and 
$\delta_{\rm c\omega}\to 3/2$ when $\Omega_0\to 0$.  



The parametric solution is rather cumbersome.  
It happens that the relation between $\delta_0$ and $\Delta$ is 
rather well approximated by 
\begin{equation}
 1+\Delta \approx (1-\delta_0/\delta_{\rm c\omega})^{-\delta_{\rm c\omega}}.
 \label{approx}
\end{equation}
Similarly, it is also useful to have an approximation to the exact 
solution for the linear theory growth factor.  When $\Omega_0 \le 1$, 
then the linear theory growth factor is well approximated by 
\begin{equation}
 D(t) \approx {(5/2)\, a(t)\,\Omega(t)\over \Omega(t)^{4/7} + 1 + \Omega(t)/2},
 \label{linthy}
\end{equation}
\cite{cpt92}, 
where $a(t)$ denotes the expansion factor at time $t$, and 
$\Omega(t)$ is given by equation~(\ref{Omegat}).  
This expression is normalized so that $D(t_0) = a(t_0) = 1$ if $\Omega_0=1$.


\subsection{Environment and spherical evolution}
Suppose we consider the evolution of a spherical underdense region 
in an $\Omega_0=1$ universe.  Let $1+\Delta_\omega < 1$ denote the 
density in this region.  If we wish to think of this region as being 
an underdense universe, then the effective value of $\Omega$ in this 
region is smaller than unity for two reasons:  first, because the 
density is lower by a factor of $1+\Delta_\omega$, and second because 
the region is expanding faster than the background, so it has an 
effective Hubble constant $H_\omega$ which is larger.  

To see what equation~(\ref{scmodel}) implies for the evolution, 
let $1+\Delta_1$ denote the density of a small patch respect to the 
background density (the subscript unity denotes the fact that this 
is the overdensity with respect to a background which has critical 
density: $\Omega_0=1$).  
Now, suppose that this patch is surrounded by a region $U$ within 
which the average density is $1+\Delta_\omega$ with respect to the 
true background.  Then the smaller patch has overdensity
 $(1+\Delta_1)/(1+\Delta_\omega)$ with respect to its local background.  
If we wish to describe the local environment as has having its own 
effective cosmological parameters, then the local value of the 
Hubble constant $H_\omega$ differs from the global one $H_0$:
 $H_\omega/H_0 = g(\omega)$.  
Thus, the expressions in equation~(\ref{scmodel}) are really the 
statements that 
\begin{equation}
 1+\Delta = {1+\Delta_1\over 1+\Delta_\omega} \qquad{\rm and}\qquad
 {v_{\rm pec}\over H_\omega R} = 
 {v_{\rm pec1} - u_{\rm pec1}\over u_{\rm pec1}},
\end{equation}
where $u_{\rm pec1}$ is the peculiar velocity of the shell $U$ 
with respect to the background, had the mass within $U$ been 
smoothly distributed (we know it is not because the central region 
has density $1+\Delta_1$).  
Now, the local value of $\Omega_\omega$ within $U$ differs from 
the global value $\Omega_0=1$ both because $\Delta_\omega\ne 0$ 
and because the different expansion rate means that the local value 
of the critical density is different: 
\begin{eqnarray}
 \Omega_\omega(t_0) = {\Omega_0\,(1+\Delta_\omega)\over (H_\omega/H_0)^2} 
     = {f(\omega)\over g(\omega)^2},
\end{eqnarray}
where we have used the fact that $\Omega_0=1$.
Notice that this relation between $\Omega_\omega$ and $\omega$ is 
the same as equation~(\ref{Uomega}).  In other words, we get the 
same description for the evolution of the small scale patch if 
we treat it as having overdensity $1+\Delta_1$ with respect to 
the $\Omega_0=1$ background within which the Hubble constant is 
$H_0$, as if we describe it with respect to the local cosmological 
model $\Omega_\omega$ and $H_\omega$, and we rescale our 
definitions of density and peculiar velocity accordingly.  In addition,
using the exact expression for the age of the universe given above, 
we can see that these definitions also guarantee that $t_0$ is the same
in the both the background and the local cosmological model.

If we write the linear theory overdensity associated with 
$1+\Delta_\omega$ as  
\begin{equation}
 \delta_\omega = h(\omega),
\end{equation}
then 
\begin{equation}
 \delta_0 = {\delta_{\rm min}\over -\delta_\omega}
     \Bigl[h(\theta)-\delta_\omega\Bigr]\nonumber
 =  {\delta_{\rm c\omega}\over \delta_{\rm c1} - \delta_\omega}
     \Bigl[h(\theta) - \delta_\omega\Bigr].
\end{equation}
The term in square brackets is simply the difference in linear 
theory values for the background cosmology.  If we think of this as 
an effective linear theory overdensity in the effective cosmology, 
then the prefactor is the effective linear theory growth factor.  
It is straightforward to verify that, indeed, 
\begin{equation}
 {\delta_{\rm c\omega}/\delta_{\rm c1}\over 
  1 - \delta_\omega/\delta_{\rm c1}} = {{D_\omega}\over D_1} 
  \qquad {\rm or}\qquad
 {\delta_{\rm c\omega}\over D_\omega} = 
 {{\delta_{\rm c1} - \delta_\omega}\over D_1}.
\end{equation}
where $D_1$ is the growth factor in the background cosmology, and 
$D_\omega$ is the growth factor in the patch, \emph{at time $t_0$}. 
This last point is important, as the expansion factor $a(t_0)$ in the 
patch cosmology is not equal to the expansion factor in the background 
cosmology. In particular, we know that 
 $a_\omega(t_0)/a_1(t_0)=(1+\Delta_\omega)^{-1/3}$.  
For completeness, we note that 
\begin{equation}
 \delta_{cw} = \delta_{\rm min} \,
   \left(1 + \frac{(2\pi)^{2/3}}{(\sinh\omega - \omega)^{2/3}}\right)
\end{equation}
(recall that we are in an underdense region).  

In the following, take $a_1(t_0)=1$, so $D_1=1$. 
The approximate solution~(\ref{approx}) of the spherical evolution 
model shows similar behaviour:  
\begin{eqnarray}
 1 + \Delta &\equiv& {1 + \Delta_1\over 1 + \Delta_\omega} 
   = \left({1-\delta_1/\delta_{\rm c1}\over 
         1-\delta_\omega/\delta_{\rm c1}}\right)^{-\delta_{\rm c1}}
  = \left({\delta_{\rm c1}-\delta_1\over 
         \delta_{\rm c1}-\delta_\omega}\right)^{-\delta_{\rm c1}}\nonumber\\
  &=& \left(1 - {\delta_1-\delta_\omega\over 
             \delta_{\rm c1}-\delta_\omega}\right)^{-\delta_{\rm c1}}
  = \left(1 - D_\omega {\delta_1-\delta_\omega\over\delta_{{\rm c}\omega}}
    \right)^{-\delta_{\rm c1}}\nonumber\\
  &\approx& 
   \left(1 - {\delta_1-\delta_\omega\over\delta_{{\rm c}\omega}/D_\omega}
    \right)^{-\delta_{{\rm c}\omega}},
\end{eqnarray}
where $\delta_1$ denotes the linear theory value associated with 
the nonlinear density $\Delta_1$ for $\Omega_0=1$.  
The final approximation follows from recalling that 
$\delta_{{\rm c}\omega}$ depends only weakly on cosmology.  
Comparison with equation~(\ref{approx}) shows explicitly that 
the relevant linear theory quantity is the difference between the 
$\Omega_0=1$ values for the perturbation and the environment, and 
this difference must be multiplied by the linear growth factor 
$D_\omega$ in the effective cosmology.  

Now, to estimate the mass function of virialized objects, we are 
interested in the case when $\theta=2\pi$.  The analysis above 
shows that 
 $\delta_{\rm c\omega}/D_\omega = \delta_{\rm c1} - \delta_\omega$;
the objects which form in a region of nonlinear density 
$1+\Delta_\omega$ with respect to the background, with corresponding 
linear overdensity $\delta_\omega$, can either be thought of as 
forming in an effective $\Omega_\omega$ cosmology \cite[e.g.,][]{gv04}, 
or as forming in the true 
$\Omega_0$ background cosmology but with an effective linear theory 
overdensity which is offset by $\delta_\omega$ to account for the 
surrounding overdensity \cite[e.g.,][]{mw96, st02}.  
The second description is easier to implement, and follows naturally 
from the excursion set description.  In particular, the analysis above 
shows that approaches which do not correctly compute $\Omega_\omega$ 
(e.g., Gottl\"ober et al. 2003 ignore the fact that $H_\omega\ne H_0$) 
are incompatible with the excursion set approach.  
In any case, the analysis above suggests that such approaches are 
ill-motivated.

\section{Formation histories}\label{evolution}
The previous section showed that the excursion set approach 
results in the same expressions for the environmental dependence 
of the present day linear theory growth factor as one derives 
from thinking of the environment as defining an effective 
cosmology.  So the question arises as to whether or not the 
two approaches predict the same evolution.  For example, one 
might have wondered if the formation histories of objects are 
the same in these two approaches.  

To see that they are, it will be convenient to modify our notation 
slightly.  We showed that 
\begin{equation}
 \frac{\delta_{\rm c}(\Omega_{\omega0})}{D_{\omega 0}} 
 = \frac{\delta_{\rm c}(\Omega_{0}) - \delta_{\rm L}(\Delta_{0})}{D_0}
 \label{lowz}
\end{equation}
where the subscripts 0 mean the present time.  
The quantity $\delta_{\rm L}(\Delta_{0})$ is what we previously 
called $\delta_\omega$; it is the value of the initial overdensity 
extrapolated using linear theory (of the background cosmology) to 
the time at which the nonlinear density is $\Delta_0$.  
Also, we previously had set the growth factor in the background 
universe at the present time to unity:  $D_0=1$.  
We have written it explicitly here to show that, had we chosen to 
perform the calculation for some earlier time, then we would have 
found 
\begin{equation}
 \frac{\delta_{\rm c}(\Omega_{\omega1})}{D_{\omega 1}} 
 = \frac{\delta_{\rm c}(\Omega_{1}) - \delta_{\rm L}(\Delta_{1})}{D_1},
 \label{highz}
\end{equation}
where the subscript 1 denotes the earlier time.  
I.e., $\Omega_{\omega 1}$ is the effective cosmology associated 
with the overdensity $\Delta_1$, which itself is related to 
$\Delta_0$ by the spherical evolution model (the region that is 
$\Delta_0$ today was a different volume in the past, but its mass 
was the same.)  And, analogously to the previous expression,
 $\delta_{\rm L}(\Delta_1)$ is the initial overdensity extrapolated 
using linear theory to the (earlier) time at which the nonlinear 
density was $\Delta_1$.  Since $\Delta_1$ is closer to 0 than is 
$\Delta_1$, $\delta_{\rm L}(\Delta_1)$ is also closer to 
0 than is $\delta_{\rm L}(\Delta_0)$.

If one were to apply the excursion set approach to study formation 
histories in the effective cosmology, one would be interested in 
the difference between equations~(\ref{highz}) and~(\ref{lowz}):
\begin{eqnarray}
 \frac{\delta_{\rm c}(\Omega_{\omega1})}{D_{\omega 1}} -
 \frac{\delta_{\rm c}(\Omega_{\omega0})}{D_{\omega 0}} &=& 
 \frac{\delta_{\rm c}(\Omega_{1})}{D_1}-\frac{\delta_{\rm c}(\Omega_{0})}{D_0} 
 \nonumber\\
 &&\ - \left[\frac{\delta_{\rm L}(\Delta_{1})}{D_1} 
             - \frac{\delta_{\rm L}(\Delta_{0})}{D_0}\right].
\end{eqnarray}
Now, the quantity in square brackets is 
\begin{equation}
 \frac{\delta_{\rm L}(\Delta_{1})}{D_1} 
       - \frac{\delta_{\rm L}(\Delta_{0})}{D_0} = 
\left[\frac{\delta_{\rm L}(\Delta_{1})}{D_1/D_0} 
       - \delta_{\rm L}(\Delta_{0})\right] \,  D_0^{-1} = 0,
\end{equation}
because $\delta_{\rm L}(\Delta_1)$ and $\delta_{\rm L}(\Delta_0)$ are 
the same quantity (the initial overdensity), evolved using linear 
theory to two different times.  In particular, $\delta_{\rm L}(\Delta_1)$ 
is closer to 0 than is $\delta_{\rm L}(\Delta_0)$ by 
 $\delta_{\rm L}(\Delta_1)/\delta_{\rm L}(\Delta_0) = D_1/D_0$.  
Thus, 
\begin{equation}
 \frac{\delta_{\rm c}(\Omega_{\omega1})}{D_{\omega 1}} -
 \frac{\delta_{\rm c}(\Omega_{\omega0})}{D_{\omega 0}} 
  = \left[\frac{\delta_{\rm c}(\Omega_{1})}{D_1/D_0} - 
           \delta_{\rm c}(\Omega_{0})\right] \, D_0^{-1} .
\end{equation}
Note that the expression on the right has {\em no} dependence on 
the effective cosmology.  Moreover, it is exactly the same as the 
expression that one obtains when using the excursion set approach 
to study formation histories in the background cosmology.  
It is in this sense that the formation histories of objects are 
independent of the effective cosmology of the environment;
the excursion set approach is a simple self-consistent way of 
exploiting this fact.  


\section{Discussion}
The excursion set description provides a simple, self-consistent 
way of estimating the effect of environment on structure formation 
and evolution.  
In particular, it is equivalent to using the fact that the large 
scale environment can be thought of as providing an effective 
background cosmology of the same age (Section~\ref{equivalence}).  
Estimating the parameters of the effective cosmology is slightly 
more involved, but useful for running simulations which mimic 
the formation of structure in different environments.  

In essence, the equivalence between the excursion set and effective 
cosmology descriptions is a consequence of Birkhoff's theorem:  
the evolution of a perturbation does not depend on its surroundings.  
There has been recent interest in models with modified gravitational 
force laws \cite[e.g.,][]{ssys05, fritz06, sshsy07}.
Since Birkhoff's theorem does not apply in such models \cite{mss08, sk2008, dai2008, cli2006, cap2007}, it will be interesting to see if this equivalence survives.  
If not, the enviromental dependence of clustering may be added as 
another constraint on such models.  

\section*{Acknowledgements}
RKS thanks Bepi Tormen for asking about this equivalence on more 
than one occasion, and the participants of the meeting on 
Cosmological Voids held in December 2006 at 
the Royal Netherlands Academy of Arts and Sciences.  
We also thank E. Neistein for insisting that the excursion set 
and effective cosmology approaches could not be reconciled with 
one another.

\label{lastpage}

\end{document}